\documentclass[twocolumn,prl,preprintnumbers,superscriptaddress,showpacs]{revtex4}

\usepackage{graphicx}
\usepackage{amsmath,amsfonts,amssymb,amsxtra}
\usepackage{units}

\makeatletter
\DeclareRobustCommand{\chemical}[1]{%
  {\(\m@th
   \edef\resetfontdimens{\noexpand\)%
       \fontdimen16\textfont2=\the\fontdimen16\textfont2
       \fontdimen17\textfont2=\the\fontdimen17\textfont2\relax}%
   \fontdimen16\textfont2=2.7pt \fontdimen17\textfont2=2.7pt
   \mathrm{#1}%
   \resetfontdimens}}
\DeclareRobustCommand{\bchemical}[1]{%
  {\(\m@th
   \edef\resetfontdimens{\noexpand\)%
       \fontdimen16\textfont2=\the\fontdimen16\textfont2
       \fontdimen17\textfont2=\the\fontdimen17\textfont2\relax}%
   \fontdimen16\textfont2=2.7pt \fontdimen17\textfont2=2.7pt
   \mathbf{#1}%
   \resetfontdimens}}
\makeatother

\newcommand{\nsmovz}{\chemical{Nd_{0.33}Sr_{1.67}MnO_4}}

\newcommand{\pcmovz}{\chemical{Pr_{0.33}Ca_{1.67}MnO_4}}

\newcommand{\lsmohd}{\chemical{La_{0.5}Sr_{1.5}MnO_4}}
\newcommand{\lsmovo}{\chemical{La_{0.42}Sr_{1.58}MnO_4}}
\newcommand{\lsmhd}{\chemical{La_{0.5}Sr_{1.5}MnO_4}}

\newcommand{\lscoz}{\chemical{La_{1.67}Sr_{0.33}CoO_4}}
\newcommand{\lscov}{\chemical{La_{2-x}Sr_{x}CoO_4}}

\newcommand{\tn}{T$_\text{N}$}

\newcommand{\kmv}{\bm k$_{{Mn^{4+}}}$}
\newcommand{\kmd}{\bm k$_{{Mn^{3+}}}$}

\newcommand{\Q}{\bm Q}

\newcommand{\mnd}{Mn$^{3+}$}
\newcommand{\mnv}{Mn$^{4+}$}
\begin{document}

\textheight 24.8 true cm

\title{ Hourglass dispersion in overdoped single-layered manganites}

\author{H. Ulbrich}
\email{ulbrich@ph2.uni-koeln.de}%
\affiliation{II.~Physikalisches Institut, Universit\"at zu K\"oln, Z\"ulpicher Str.~77, D-50937 K\"oln, Germany}

\author{P. Steffens}
\affiliation{Institut Laue-Langevin, BP 156, 38042 Grenoble Cedex
9, France}

\author{D. Lamago}
\affiliation{Forschungszentrum Karlsruhe, Institut f\"ur
Festk\"orperphysik, P.O.B. 3640, D-76021 Karlsruhe, Germany}
\affiliation{Laboratoire L\'eon Brillouin, C.E.A./C.N.R.S.,
F-91191 Gif-sur-Yvette Cedex, France}

\author{Y. Sidis}
\affiliation{Laboratoire L\'eon Brillouin, C.E.A./C.N.R.S.,
F-91191 Gif-sur-Yvette Cedex, France}

\author{M. Braden}
\email{braden@ph2.uni-koeln.de} \affiliation{II.~Physikalisches Institut, Universit\"at zu K\"oln, Z\"ulpicher Str.~77, D-50937 K\"oln, Germany}

\date{\today, \textbf{preprint}}


\date{\today}
\begin{abstract}

The incommensurate stripe-like magnetic ordering in two
single-layered manganites, Nd$_{0.33}$Sr$_{1.67}$MnO$_4$ and
Pr$_{0.33}$Ca$_{1.67}$MnO$_4$, is found to exhibit an
hourglass-like excitation spectrum very similar to that seen in
various cuprates superconductors, but only for sufficiently short
correlation lengths. Several characteristic features of an
hourglass dispersion can be identified: enhancement of intensity
at the merging of the incommensurate branches, rotation of the
intensity maxima with higher energy transfer, and suppression of
the outwards-dispersing branches at low energy. The correlation
length of the magnetic ordering is identified as the decisive
parameter causing the hourglass-shape of the spectrum.

\end{abstract}

\maketitle

\begin{figure}
\includegraphics[width=0.85\columnwidth]{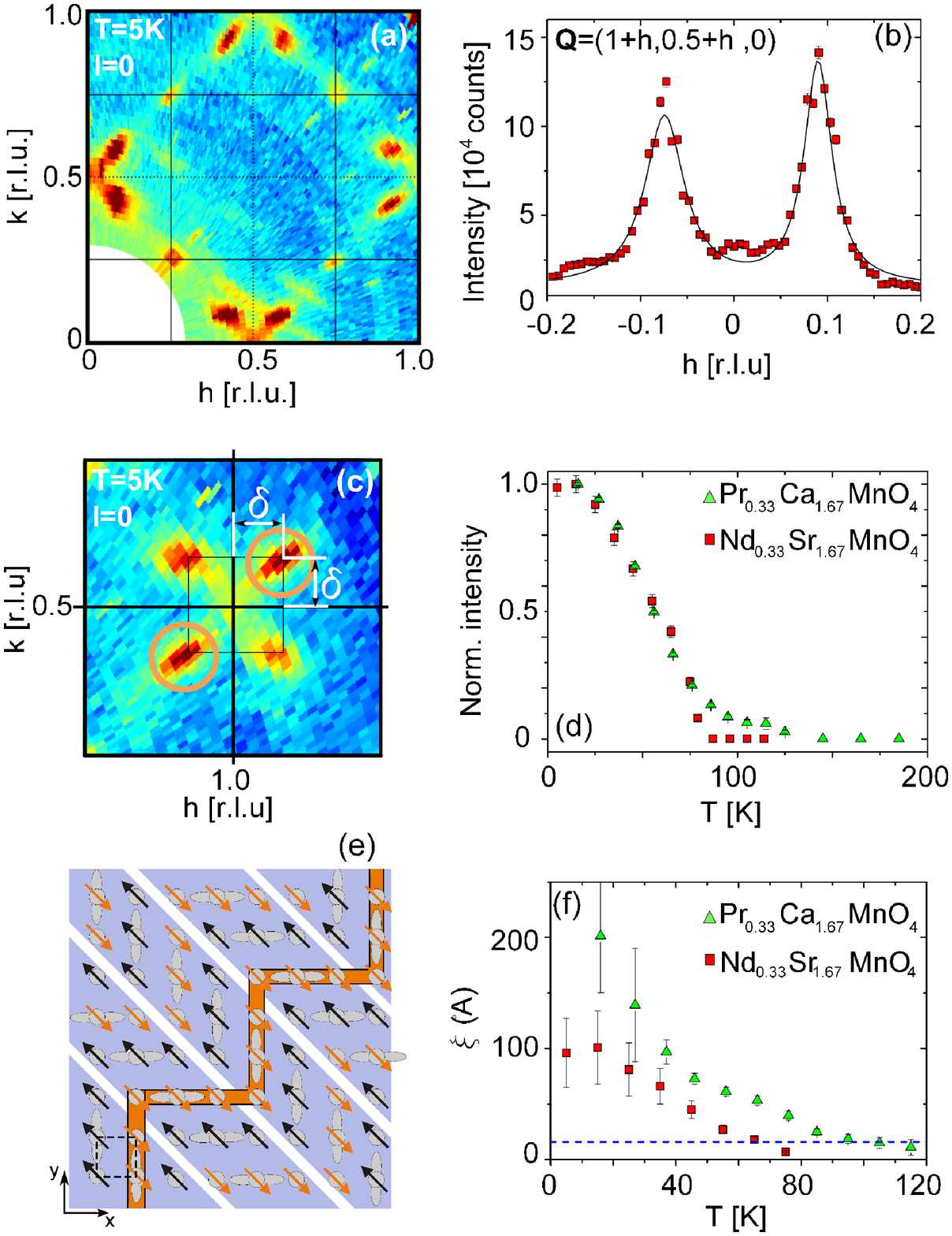}
\caption{(color online). Map of magnetic scattering in the (hk0)-plane in \nsmovz \ at T=5K (a) and enlargement of the map (c) near {\bf
Q}=(1.0,0.5). Strong magnetic superstructure reflections are found near or at the positions of the magnetic intensities in half-doped \lsmhd :
propagation vectors are \kmd=$\pm$(0.25,0.25,0) and \kmv=$\pm$(0.5+$\delta$,$\delta$,0) with $\delta$=0.083$\sim1/12$. (e) sketch of the orbital
and magnetic structure explaining in $2/3$-doped layered manganites; the slabs of strong magnetic correlations mediated through the orbital
ordering are indicated by the diagonal ribbons. (b) Scan across two magnetic satellites and the common commensurate position. (d) (1,0.5)Peak
height and (f) correlation length of the magnetic scattering along the modulation plotted as a function of temperature for the two compounds;
the dashed line indicates the magnetic period which amounts to twice the distance of charge and orbital stripes.} \label{elastisch}
\end{figure}

The implication of the stripe instability \cite{1} in the pairing
mechanism of high-temperature superconducting cuprates remains
matter of strong controversy. Static stripe order competes with
superconductivity in the cuprates, as superconductivity is
strongly suppressed in samples with static stripe order \cite{2}.
However fluctuating stripes may coexist with good superconducting
properties and could be important for the pairing. The shape of
the dispersion of magnetic excitations in various superconducting
cuprates \cite{3,4,5} is taken as evidence that the stripe concept
is the correct basis to understand the magnetic correlations even
in highly doped cuprates, but analyzes within a fully itinerant
picture also yield reasonable description \cite{6}. At low energy,
magnetic excitations can be associated with four incommensurate
spots close to a common commensurate propagation vector, (0.5,0.5)
\cite{note}. With increasing energy the q-position varies little
until a merging of different branches occurs at (0.5,0.5), which
is associated with an enhanced signal. In constant energy maps at
higher energy, the strongest scattering appears again at
incommensurate positions, but intensity maxima are rotated by
45$^\circ$ \ with respect to those at low energy. This shape of
the dispersion has been observed in several cuprates by now
\cite{3,4,5,7}.

\begin{figure}
\includegraphics[width=0.9\columnwidth]{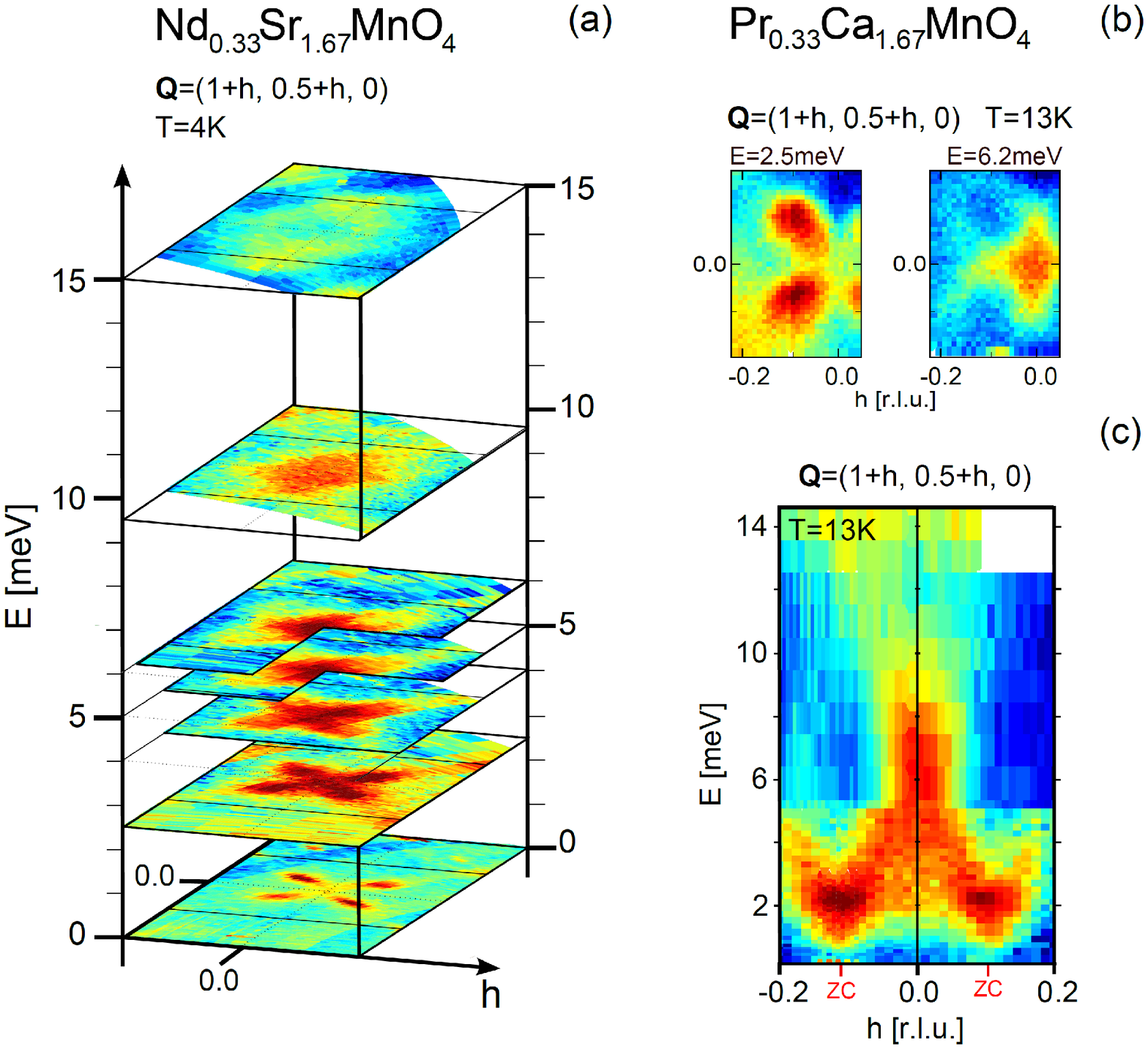}
\caption{(color online). Constant energy slices from \nsmovz \ around the four incommensurate antiferromagnetic spots \Q =(1$\pm\delta$, 0.5$\pm
\delta$,0) with $\delta$=0.083 measured at T=4K  with the IN20 flatcone spectrometer (a). Two constant energy slices from \pcmovz \ around the
same incommensurate antiferromagnetic zone-centers \Q =(1$\pm\delta$, 0.5$\pm \delta$,0) measured at T=13K with the 1T spectrometer (b). Part
(c) shows an energy versus Q map with the chosen Q-direction passing across two incommensurate and the central commensurate positions. Data
taken on the cold 4F.2 spectrometer and on the thermal 1T spectrometer were combined for this plot. } \label{3dimplot}
\end{figure}

Stripe phases were reported in other layered transition-metal
oxides \cite{8,9}. Those appearing in isostructural nickelates
\cite{8} have been studied most, but the dispersion of magnetic
excitations in these materials does not show the characteristic
features of the hourglass dispersion \cite{10} most likely due to
the remarkable stability  and particular ratio of interactions
parameters of the magnetic order at $1/3$ doping \cite{10,12}.
More recently stripe-like magnetic ordering was also found in the
layered isostructural cobaltates, \lscov \cite{9}, and the
magnetic excitations in one of these compounds, \lscoz , were
found to exhibit a dispersion very similar to the hourglass
dispersion in the cuprates \cite{12}.

Stripe phases have been reported also for manganites with a doping level above half doping \cite{13,14,15}. In addition to the coupling of
charges and magnetic moments in cuprates and in nickelates, the orbital degrees of freedom are decisive in the manganites. Only recently a
conclusive model of stripe-like magnetic and electronic ordering in manganites was elaborated by the aid of comprehensive neutron diffraction
experiments \cite{16}. In this work we analyze the magnetic excitations in two overdoped manganites with incommensurate magnetic order,
Nd$_{0.33}$Sr$_{1.67}$MnO$_4$ and Pr$_{0.33}$Ca$_{1.67}$MnO$_4$. Both compounds exhibit the characteristic features of the hourglass dispersion:
the enhancement of signal at the merging of incommensurate branches, the rotation of the incommensurate intensity distribution with increasing
energy and the suppression of the outwards dispersing signals. Most importantly, we find that the hourglass shape of the dispersion only appears
for short magnetic correlation lengths.

Two large single-crystals of Nd$_{0.33}$Sr$_{1.67}$MnO$_4$ and
Pr$_{0.33}$Ca$_{1.67}$MnO$_4$ have been grown by a floating-zone
image furnace \cite{17}. The quality of the samples has been
checked by macroscopic and by microscopic investigations. Both
systems exhibit a transition into a charge and orbital ordered
state and, at lower temperatures, a transition into an
antiferromagnetic state. The magnetic susceptibility of \pcmovz \
measured in a SQUID magnetometer exhibits a clear kink at the
transition into the charge and orbital ordered state at T=320K
which is in perfect agreement with published data, \cite{18}
whereas the charge and orbital order in
Nd$_{0.33}$Sr$_{1.67}$MnO$_4$ is not visible in the
susceptibility. The transitions into the antiferromagnetic state
have been characterized by neutron diffraction on both compounds
(Fig.1). Both systems exhibit nearly the same transition
temperatures (\tn =100K for \pcmovz and \tn = 80K for \nsmovz ),
but the order is better defined in \pcmovz \ which exhibits
sharper magnetic superstructure peaks. The two compounds posses
the same amount of electronic doping, but due to the smaller ionic
radius of Ca$^{2+}$ \pcmovz \ exhibits a structural distortion
associated with the rotation of the MnO$_6$ octahedra.
Neutron-scattering experiments were performed with the thermal
spectrometers IN20 (k$_f$=3.00\AA$^{-1}$) and IN3
(k$_f$=2.66\AA$^{-1}$) at the ILL and 1T (k$_f$=2.66\AA$^{-1}$) at
the Orphe\'ee reactor. In addition, the cold triple-axis
instrument 4F.2 (k$_f$=1.55\AA$^{-1}$) at LLB was used to study
magnetic excitations at low energies.

\begin{figure*}
\centering
\includegraphics[width=0.94\textwidth]{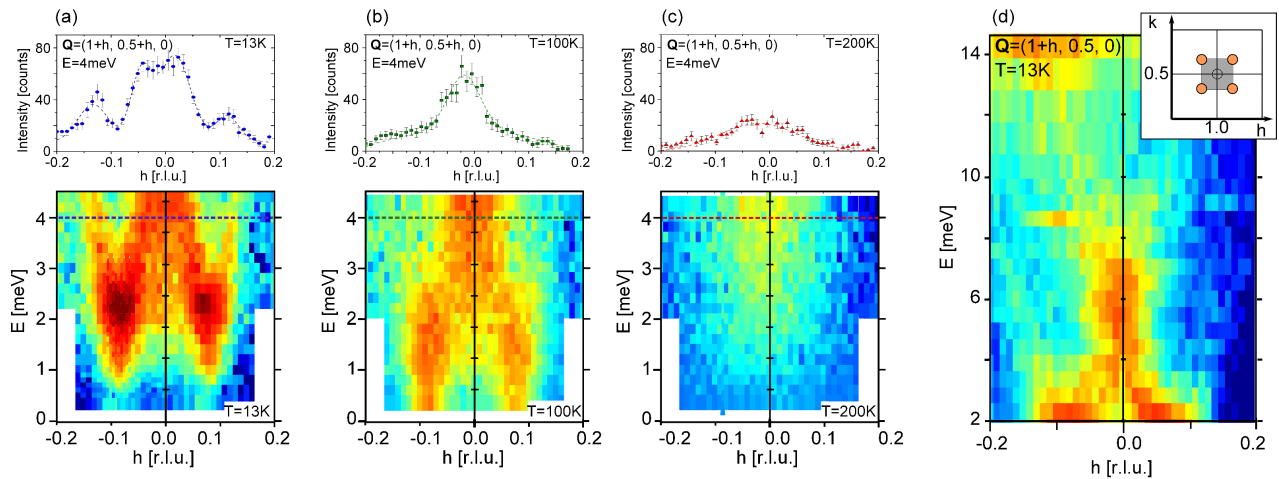}
\caption{(color online). [Lower panels (a-c)] energy-versus-$Q$ maps of the magnetic excitation spectrum in \pcmovz \ at T=13K (a), T=100K (b)
and T=200K (c) taken on the cold-source spectrometer 4F.2; the chosen Q-direction passes through the commensurate magnetic position (1.0,0.5)
and two incommensurate magnetic zone centers. [Upper panels (a-c)], constant energy cuts along the same $Q$ path for an energy transfer of 4\
meV; intensities were corrected for the Bose factor to simplify comparison. Map of magnetic excitations (Q versus energy) taken on the thermal
spectrometer 1T (d); the chosen Q-direction passes through the commensurate magnetic position (0.5,0.5) along \Q =(1+h, 0.5, 0)}
\label{tempabh}
\end{figure*}

Figure 1 shows a sketch of the magnetic order and a map of elastic
scattering obtained in \nsmovz \ at low temperature. For \pcmovz \
we find very similar elastic patterns indicating essentially the
same magnetic ordering. By comparison with the various types of
order identified in slightly overdoped \lsmovo \ \cite{16} the
different ordering patterns can be easily understood. The
incommensurate superstructure intensities appearing at large
Q-vector can be identified with the orbital and charge order. In
addition, there are two types of magnetic signals appearing at low
Q-values. The magnetic elastic scattering in \pcmovz \ and in
\nsmovz \ differs from that in \lsmovo \ as the incommensurate and
commensurate scattering seems to be interchanged. In \pcmovz \ and
in \nsmovz \ we find incommensurate signals close to the
commensurate propagation vector of the order of \mnv \ moments in
\lsmohd \cite{19,16}, $Q$=(1$\pm\delta$, 0.5$\pm \delta$) with
$\delta$=0.083. Four satellites are thus surrounding the
commensurate center (1,0.5) similar to the arrangement of the
incommensurate magnetic signal around $Q_{AFM}$=(0.5,0.5) in the
cuprates. In addition, \pcmovz \ and \nsmovz \ show commensurate
signals exactly at the q-position of the \mnd \ spin order  in
\lsmohd \ \cite{19}. Like in the case of \lsmovo , the magnetic
signals can be explained assuming that additional \mnv \ sites
align in stripes along the diagonals, see the sketch in Fig. 1
(e). The magnetic order consists of an antiferromagnetic stacking
of ferromagnetic zigzag chains similar to that in CE-type order at
half doping. However, in contrast to the ideal CE-type order the
legs of these ferromagnetic zigzag chains are four - and not three
- units long.

Maps of inelastic scattering arising from the magnetic excitations
are presented in Fig. 2 (a) and (b) for  \nsmovz \ and \pcmovz ,
respectively. The (h,k) maps taken at constant energy transfer on
\nsmovz \ show that the elastic scattering and that at the low
energy transfer of 2.5meV look qualitatively the same. At slightly
higher energy, the incommensurate signals shift towards the
commensurate position where a single peak appears in the 6meV
energy map. Further increase of the energy transfer results in a
splitting of the signal. However, these intensity maxima appear
now along the axes of the (h,k) plane and are thus rotated by
45$^\circ$ with respect to the low-energy ones. Fig. 2 (b) shows a
map containing the energy and the $Q$-path cutting through two
satellites and the commensurate center and energy cuts below and
above the spin-wave merging. Fig. 3 shows the same maps at
different temperatures and Fig. 4 an energy-versus-$Q$ map for the
$Q$-path running through the commensurate center (0.5,0.5) along
the axis. Also this latter map shows the merging of spin-wave
branches starting at the incommensurate satellites into the
center. \pcmovz \ exhibits qualitatively the same dispersion as
\nsmovz , but we already want to emphasize that the \pcmovz \ map
taken at low temperature also exhibits outwards dispersing
branches. Only at higher temperature, when the correlation length
has become shorter, these outwards branches are suppressed giving
rise to the full hourglass shape. The shape of the dispersion in
both these insulating materials strongly resembles the hourglass
distribution in the cuprates \cite{4,5} as it was also reported
for \lscov \ \cite{12}. This is most visible in the enhancement of
the signal towards the commensurate center and in the rotation of
the position of the satellites with increasing energy. The
hourglass dispersion can thus be considered as the normal result
of a stripe arrangement with short correlation length, and the
nickelates case appears special most likely due to the particular
stability of this magnetic phase \cite{10}. In the manganites the
scattering at the incommensurate satellites exhibits a clear gap
due to single-ion anisotropy similar to behavior in the half-doped
compounds \cite{19,20} rendering this dispersion even more similar
to that in the cuprates.

The similarities between the hourglass shape of magnetic
excitations in metallic cuprates and the dispersion arising from
an incommensurate insulating stripe phase is remarkable, but it
does not fully exclude other explanations of the magnetic response
in the cuprates \cite{6}. In particular the resonance mode in the
superconducting phase in cuprates requires further attention.

We have also studied the magnetic excitations in \lsmovo \ finding
again an hourglass type of dispersion at low energies. Due to the
much smaller pitch of the incommensurate modulation in \lsmovo \
the merging of the spin-wave branches appears at much lower
energy, so that a high energy resolution is required strongly
limiting the statistics.

Some of the features of the hourglass dispersion can be well
described within linear spin-wave theory \cite{21}. For example
the merging of the branches and the rotation of the incommensurate
scattering in constant energy cuts above the merging are just
consequences of the dispersion of the branches starting at
different satellites. However spin-wave theory predicts that both
the inwards and the outwards dispersing branches starting from the
satellites posses considerable intensity. In contrast, the
hourglass spectrum gains its characteristic shape by the
suppression of the outwards dispersing branches. In the magnetic
response of the cuprates there is no indication of these branches
left. In this aspect the low-temperature spectrum in \pcmovz \
looks different, as one can clearly identify the outwards
dispersing branches as well. These outwards modes present a
significant difference to the hourglass spectrum in the cuprates.
By varying the temperature, \pcmovz \ offers the possibility to
analyze the origin of this important aspect of the hourglass
spectrum. At the intermediate temperature of 100\ K the outwards
modes are severely suppressed, so that the magnetic response of
\pcmovz \ becomes fully similar to the hourglass spectrum. Further
increase of the temperature up to 200\ K reduces the signs of the
incommensurate scattering in this inelastic energy-versus-$Q$ map.
In \pcmovz , the suppression of the outwards branches clearly
follows the reduction of the correlation length perpendicular to
the stripes shown in Fig. 1(f). In \nsmovz \ the correlation
length of magnetic ordering across the stripes remains reduced
even at the lowest temperature in accordance with a pronounced
hourglass shape persisting. We may thus conclude that the
correlation length of the magnetic stripe order has an essential
impact on the shape of the magnetic response. For long correlation
lengths, as present at low temperature in \pcmovz , the outwards
and the inwards dispersing branches posses comparable weight,
whereas the response focusses into the inwards dispersing branches
for shorter correlation lengths.

When the hourglass-shaped dispersion is observed, both manganites exhibit very short longitudinal magnetic correlation lengths, of the order of
or even below the length of the magnetic period (i.e. twice the stripes distance), whereas the correlation of the orbital order is still
sizeable. A smoothly decaying magnetic order parameter is no longer appropriate to describe such short range magnetic ordering, but one may
conclude that magnetically correlated regions cover only a few charge stripes in these states in spite of a well defined modulation direction.
This is reminiscent of a nematic state \cite{23}.

The reason for the short correlation length in the two $2/3$-doped
manganites can be understood by inspecting Fig. 1(e). Orbital
ordering forms slabs running along the diagonals. In these
three-Mn thick slabs the magnetic order is well determined: the
orientation of the ordered $e_g$ orbital couples the moments of
the two \mnv \ ferromagnetically with the central \mnd \ moment;
neighboring \mnv -\mnd -\mnv \ trimers are coupled
antiferromagnetically, so that all moments within a slab are well
fixed by magnetic interaction parameters that can be expected
being very similar to those in \lsmohd \cite{24}. The slabs are
indicated as shaded areas in Fig. 1 (e). However, the interaction
between two slabs is frustrated in the ideal tetragonal structure,
and some additional distortion is needed to stabilize the four-leg
zigzag structure. The weak inter-slab interaction causes the rapid
melting of magnetic order much below the suppression of orbital
ordering. It is the state with loosely coupled magnetic slabs,
which exhibits the hourglass dispersion. This situation is very
similar to the short-range magnetic order in \lscoz \ \cite{12,9}
and in cuprates. In both these cases magnetic coupling across the
stripes is very weak compared to the nearest-neighbor coupling
active within the antiferromagnetic regions. The hourglass
spectrum in the cuprates was analyzed in various models for
itinerant or local moments with or without static order
\cite{22,22a,22b}. The model of weakly coupled ladders corresponds
to a very short correlation length of only one ladder size and
results in the suppression of the outwards modes \cite{22}. In
contrast, the strong coupling across the charge stripe in the
nickelates results in more stable magnetic order \cite{10}
reflected by an isotropic spin-wave dispersion.

In conclusion the dispersion of magnetic excitations in strongly overdoped layered manganites closely resembles the hourglass spectrum seen in
the cuprates and in a layered cobaltate. Some of the characteristic hourglass features can be attributed to the incommensurate modulation of the
magnetism in these stripe phases. These are the enhanced signal at the merging point of the branches and the rotation of the incommensurate
signal in constant energy maps. However, the suppression of the outwards-dispersing branches, which is the essential element of the hourglass
spectrum, can be attributed to the reduction of the correlation lengths. In \pcmovz , we find significant intensity in the outwards dispersing
branches at low temperature, where the correlation lengths are long. These branches, however, become suppressed upon heating and decrease of the
correlation lengths, giving rise to the hourglass spectrum. In the cuprates, in the cobaltate and in the manganites studied here, the hourglass
dispersion thus appears in a state with slabs of well defined magnetic correlations, which are only loosely coupled.

This work was supported by the Deutsche Forschungsgemeinschaft through the Sonderforschungsbereich 608.


\begin{thebibliography}{}
\bibitem{note} Throughout this paper, we use
two-dimensional notations neglecting the weak inter-layer coupling
($a\sim3.8$\AA ).
\end{thebibliography}
\end{document}